\documentclass[fleqn,usenatbib]{mnras}
\usepackage[utf8]{inputenc}  
\usepackage{newtxtext,newtxmath}

\usepackage[T1]{fontenc}

\DeclareRobustCommand{\VAN}[3]{#2}
\let\VANthebibliography\thebibliography
\def\thebibliography{\DeclareRobustCommand{\VAN}[3]{##3}\VANthebibliography}

\usepackage{graphicx}	
\usepackage{amsmath}	
\usepackage{amssymb}	

\usepackage[dvipsnames]{xcolor}
\usepackage{lastpage}

\title[Using ANNs to detect OTs with images from different surveys]{Detecting optical transients using artificial neural networks and reference images from different surveys}

\author[K. Wardęga et al.]{
Katarzyna Wardęga,$^{1,2}$\thanks{E-mail: katarzyna.wardega96@gmail.com}
Adam Zadrożny,$^{3}$\thanks{E-mail: adam.zadrozny@ncbj.gov.pl}
Martin Beroiz,$^{1}$
Richard Camuccio $^{1,4}$\newauthor
and Mario C. D\'{\i}az$^{1}$
\\
$^{1}$Center for Gravitational Wave Astronomy, The University of Texas Rio Grande Valley, Brownsville, TX 78520, USA\\
$^{2}$Faculty of Physics, University of Warsaw, Poland\\
$^{3}$National Centre for Nuclear Research, Astrophysics Division, ul. Ho\.{z}a 69, 00-681 Warsaw, Poland\\
$^{4}$Department of Physics and Astronomy, Texas Tech University, Lubbock, TX 79409, USA
}

\date{Accepted XXX. Received YYY; in original form ZZZ}

\pubyear{2020}

\begin{document}
\label{firstpage}
\pagerange{\pageref{firstpage}--\pageref{lastpage}}
\maketitle

\begin{abstract}
To search for optical counterparts to gravitational waves, it is crucial to develop an efficient follow-up method that allows for both a quick telescopic scan of the event localization region and search through the resulting image data for plausible optical transients. We present a method to detect these transients based on an artificial neural network. We describe the architecture of two networks capable of comparing images of the same part of the sky taken by different telescopes. One image corresponds to the epoch in which a potential transient could exist; the other is a reference image of an earlier epoch. We use data obtained by the Dr. Cristina V. Torres Memorial Astronomical Observatory and archival reference images from the Sloan Digital Sky Survey. We trained a convolutional neural network and a dense layer network on simulated source samples and tested the trained networks on samples created from real image data. Autonomous detection methods replace the standard process of detecting transients, which is normally achieved by source extraction of a difference image followed by human inspection of the detected candidates. Replacing the human inspection component with an entirely autonomous method would allow for a rapid and automatic follow-up of interesting targets of opportunity. The method will be further tested on telescopes participating in the Transient Optical Robotic Observatory of the South Collaboration. 
\end{abstract}




\section{Introduction}

Detecting optical transients (OTs) associated with gravitational wave (GW) events is one of the challenges in time domain astronomy.
The GW localization region on the sky is typically large ($\sim$100 square degrees) when two or three GW detectors are operating \citep{abbott_2019, coughlin_2020}. When following up GW events, it is crucial for observatories with $<$1 square degree field-of-view (FOV) to develop efficient methods to detect plausibly-linked transients \citep{coughlin_2020}. Successfully detecting an electromagnetic counterpart to a GW reduces the localization uncertainty on the sky and allows for further characterizing measurements.

The primary goal of this paper is to describe a method for detecting transients by comparing two images of the same region of the sky taken at different times and by different telescopes. The method is based purely on machine learning (ML) algorithms, specifically artificial neural networks (ANNs). The ML approach to transient detection is efficient as it can search through a large data set in a short amount of time. Hence, the ML approach is a valuable approach to solving the problem of detecting OTs in the time domain.

Difference image analysis (DIA) is the standard method used to search for OTs. DIA methods are based on subtracting a reference image from a target image. The method attempts to compensate for the difference in point spread functions (PSFs) of each image. Compensating for differences in PSF allows one to subtract images taken by different telescopes or under varying atmospheric conditions. However, even with PSF compensation, the resulting image difference will be imperfect, leaving behind residual flux that can be confused by detection algorithms as false OTs. Many DIA algorithms have been proposed since the original \cite{alard_1998} paper, notably those by \cite{bramich_2008} and \cite{zackay_2016}.

Regardless of the DIA method used, it is customary to train ML agents (e.g. random forest algorithms or neural networks) to sift through all the OT candidates, remove the spurious subtraction artifacts (``bogus sources''), and retain the likeliest true OT candidates \citep{zwickyRB, iPTFRB01, ogleRB, diaz_2016, artola_2020}. A real/bogus classifier can be avoided if there is a manual operator, but it becomes cumbersome for large surveys where bogus sources can outnumber potential real ones by 100 to 1. For these reasons, most systematic searches of the sky, like those for GW optical counterparts, will require a real/bogus classifier at the end of the analysis pipeline.

Since ML classification seems like an unavoidable element in the search for OTs, we propose training ML algorithms on the images directly and avoiding the necessity of DIA methods. An ML classifier takes two small image insets which are cropped around a detected source on the target image. One inset contains the detected source and the other is cropped around the same location on the reference image. When the source appears on the target inset, but is missing on the reference inset, the classifier calls the case an OT. When there is a source present on both insets, the classifier calls the case a non-OT. Providing the classifier with a sufficient number of example OT and non-OT cases will train it to be robust at detecting all true OTs on subsequent imaging runs.

Bypassing DIA has several advantages. The neural network method we propose is robust against PSF variations across different surveys and filters. Observatories lacking an extensive reference archive benefit from a method that works regardless of the references used, as long as the sky region is covered by some comparable photometric survey. Since it is typical for DIA methods to be computationally expensive, avoiding them leads to a drastic reduction in the processing speed for pipelines and an overall simplification in their design.

To test the feasibility of our proposed method, we built and trained two ANN models --- one is a convolutional neural network (CNN) and the other is a dense layer network (DLN). The models accept target-reference inset pair samples as input and return the likelihood of the inset pair being an OT as output. We trained the models on simulated data and calculated which prediction the simulations gave on test data. The test data were created from images of galaxies obtained by the Dr. Cristina V. Torres Memorial Astronomical Observatory (CTMO) and covered by the Sloan Digital Sky Survey (SDSS) \citep{gunn_1998, gunn_2006}.

\section{Scientific Background}

Binary neutron star (BNS) or neutron star-black hole (NSBH) binary systems are the most promising astrophysical events for producing electromagnetic counterparts to GWs \citep{coughlin_2020}. Compact binary mergers are expected to produce an r-process-powered thermal transient, or a ``kilonova'' \citep{Berger_2013}. Theoretical light curves for kilonovae indicate a rapidly-fading optical and near-infrared (NIR) transient, detectable by telescopes within a week of the associated GW event. The first plausible kilonova detected was associated with the short gamma-ray burst (GRB) 130603B, which the Swift and Konus-Wind satellites detected on 3 June 2013. The short GRB lasted 0.4 seconds and was observed 12 arcminutes offset from the center of the galaxy NGC 3691 \citep{frederiks_2013}. The associated NIR transient matched the expected brightness and color of a kilonova at the time of observation, providing strong evidence for its source having been a BNS or NSBH merger.

On 17 August 2017 at 12:41:04 UTC the Advanced LIGO and Virgo detectors observed the first BNS merger. GW170817 had a signal-to-noise ratio (SNR) of 32.4 and a false alarm rate (FAR) of 1 in $8.0 \times 10^4$ years. The component masses were both calculated to be in the range $1.17-1.60M_\odot$ implying the progenitors were both likely to be neutron stars. The GW signal was localized to within 28 deg$^2$ at 90\% confidence and estimated to have a luminosity distance of (40 $\pm$ 8) Mpc \citep{maggiore_2018}. Approximately 1.7 seconds following the detection of GW170817, the Fermi Gamma-Ray Burst Monitor (Fermi-GBM) and INTEGRAL satellites detected a short GRB. The short GRB uncertainty region overlapped that of the GW, improving overall localization estimates for follow-up observations. The near-simultaneous spatial and temporal localization of GW170817 and GRB 170817A had a 1 in $5.0 \times 10^{-8}$ chance of occurring randomly and, hence, is strong evidence for the link between BNS mergers being the progenitors of short GRBs \citep{abbott_2017}. Many observatories followed up this event to search for EM counterparts \citep{abbott_2017} with the Transient Robotic Observatory of the South (TOROS) Collaboration being part of the search campaign \citep{diaz_2017}. Subsequent follow-up observations detected an optical counterpart named AT2017gfo at 11 hours post-merger at approximately 10 arcseconds offset from the core of the lenticular galaxy NGC 4993 \citep{abbott_2017}. Additionally, several teams observed both X-ray and radio emission at the position of AT2017gfo at nine and 16 days post-merger, respectively \citep{abbott_2017}.

The light curves of AT2017gfo exhibited rapid luminosity change in the ultaviolet (UV), optical, and infrared (IR) bands. An initial UV-blue peak transitioned rapidly to the red and IR bands. The rate of change for the blue bands was about two magnitudes per day. The red bands declined 0.3 days for the first 1.5 days, then the decline stopped for four days and continued to slowly decline for another eight days \citep{maggiore_2018}. The color evolution is unusual for an OT and different to any previously observed type of source. The light curves match the predicted light evolution of a ``kilonova'' --- an r-process-powered thermal transient produced by the merger of a BNS or NSBH binary system. AT2017gfo fit a kilonova model with three ejecta components, each with different masses, velocities, and opacities (see Table \ref{best fit parameters}) \citep{villar_2017}. Spectroscopic observations showed that the blue spectrum was continuous and featureless, due to line broadening of the high ejecta velocities for that component. The near-IR spectrum showed the emergence of broad spectral features, related to the radioactive decay of synthesized r-process elements at late post-merger times \cite{abbott_2017}. 

\begin{table}
    \centering
    \begin{tabular}{|c|c|c|c|}\hline
        & \textbf{blue} & \textbf{purple} & \textbf{red} \\ \hline
        $\kappa$ & 0.5 cm$^2$/g & 3 cm$^2$/g & 10 cm$^2$/g \\ \hline
        $M_{\text{ej}}$ & 0.02\(M_\odot\) & 0.047\(M_\odot\) & 0.011\(M_\odot\) \\ \hline
        $v_{\text{ej}}$ & 0.27$c$ & 0.15$c$ & 0.14$c$ \\ \hline
    \end{tabular}
    \caption{The best-fit parameters for AT2017gfo using a three-component kilonova model (lanthanide-free ``blue'', intermediate-opacity ``purple'', and lanthanide-rich ``red'' components). The fitted ejecta parameters are opacity $\kappa$, mass $M_{\text{ej}}$, and velocity $v_{\text{ej}}$. Credit: Villar et al. (2017)}
    \label{best fit parameters}
\end{table}

A kilonova is significantly different than any transient previously observed. The peak luminosity of a kilonova is predicted to be $10^{41}$ erg/s, placing it between a nova ($10^{38}$ erg/s) and a supernova ($10^{43}$ erg/s) \citep{kasliwal_2011}. Kilonovae are expected to be observable on the order of 10 days, while a nova can be observed for months and a supernova for up to a year or more. Kilonovae progenitors are BNS and NSBH mergers, whereas a nova is caused by the fusion of hydrogen on the surface of a white dwarf in a binary system, and a supernova is an explosion caused by the core collapse of a massive star. For a full overview of the history of kilonovae, theoretical models, and observations, we refer the reader to \cite{metzger_2019}. A summary of all observations taken by collaborations in the follow-up of GW170817 is given by \citet{abbott_2017}. The TOROS Collaboration contribution is further detailed in \citet{diaz_2017}. 

\section{Method}
\label{sec:methods}

To test our proposed method we ran an experiment to prove the validity of our assumptions.
The experiment consists on testing two different approaches to Machine Learning architectures based on Artificial Neural Networks. Then after training and validating them with simulated data, test them on pairs of real images. One member of the pair is from the CTMO and the reference member of the pair comes from the SDSS survey.

This section is organized as follows. In section \ref{sec:testrealdata}, we describe what kind of images from CTMO were used and how were downloaded and aligned equivalent images from SDSS. In section \ref{sec:datasets}, we present how we created the testing and training data set. In section \ref{sec:cnnarch}, we describe the architecture of the networks. Finally in section \ref{sec:metricresults} we present the results of the final metric values for our experiment.

In section \ref{sec:diaanncompare} we make another similar experiment with a set of images that has been analyzed before in search for optical transients using a DIA method (\citet{artola_2020} with a Random Forest real/bogus classifier and also with a CNN-based real/bogus classifier. This second experiment allows for a more direct comparison of the method proposed here and the more conventional one based on DIA followed before.

\subsection{Image preprocessing}
\label{sec:testrealdata}

We targeted five galaxies (Table \ref{ctmo data}) covered by SDSS using the instrumentation of CTMO. Four of the five targets were taken on February 8, 2020 UTC with the current optical configuration of CTMO, which consists of a PlaneWave Corrected Dall-Kirkham 17'' astrograph with a ProLine 16803 CCD camera. Each image is unfiltered, has 60-second exposure time, taken at $2 \times 2$ binning, and has a FOV of $80 \times 80$ arcminutes. We observed the fifth target, IC 4559, at an earlier date (2 July 2019 UTC) when CTMO had a different optical setup: the instrument used for these data was an Apogee F16M CCD camera. This image is unfiltered, taken at $2 \times 2$ binning with 300-second exposure time, and has a FOV of $50 \times 50$ arcminutes.

We used the CTMO Anaylsis Library (CAL) to bias- and dark-subtract, as well as flatfield-correct, each image \citep{camuccio_2019}. We used two-dimensional spatially-varying mesh to subtract the median background of each image. Since each target consisted of a series of exposures, we plate-solved each image and aligned them per series using their world coordinate system (WCS) header metadata. We created a median-combined stack of the aligned images per series (Note: approximate limiting magnitude SDSS is 3 sigma).

\begin{table}
    \centering
    \begin{tabular}{|c|c|c|c|}\hline
        \textbf{Object} & \textbf{RA (J2000)} & \textbf{Dec (J2000)} & \textbf{Redshift} \\ \hline
        IC 4559 & 15:35:53.51 & +25:20:28.07 & 0.0345 \\ \hline
        PGC 21547 & 07:40:29.98 & +83:47:25.88 & 0.0068 \\ \hline
        PGC 21577 & 07:41:12.48 & +42:44:57.74 & 0.0358 \\ \hline
        PGC 21708 & 07:45:07.25 & +46:04:20.72 & 0.0312 \\ \hline
        PGC 21856 & 07:48:34.63 & +44:41:17.80 & 0.0204 \\ \hline
 	\end{tabular}
 	\caption{CTMO targets}
 	\label{ctmo data}
 \end{table}

We used the \textit{SkyView} function from the \textit{astroquery} \citep{ginsburg_2019} package to download reference images from SDSS. Knowing the center coordinates and FOV of each CTMO image, we requested the SDSS reference in the \textit{g} filter with a size of $2000 \times 2000$ pixels. All SDSS images are taken from Data Release 9 (DR9) and have an exposure time of 54 seconds. We expect each image in a given pair to have different orientations. For an effective alignment solution, each pixel per picture should represent the same astronomical coordinates. To achieve image alignment, CAL employs the \textit{reproject} package from \textit{Astropy}. The \textit{reproject} package aligns the SDSS image with the CTMO image and crops it to have the same FOV.

\subsection{Creating data sets}
\label{sec:datasets}

We anticipate transient events to look like new stellar sources in the sky. We wanted to construct ML methods so that they would recognize new sources in both follow-up observations and previously-observed fields. Using the entire image as input to the neural network proved burdensome. Therefore, we created a data set with smaller images --- the data set is composed of cropped images for each source detected on the images.

We created a data set of 3370 samples from five CTMO-SDSS image pairs (hereafter the ``test data set''). Half of the samples were transients and the other half were non-transients. To train any ML model, one requires many samples ($>$10000). For this reason, we simulated a data set for the training component.

\subsubsection{Test data set}

We postulate that source extraction programs could find transient events based on the assumption that they would look like stellar sources. We built transient and non-transient samples from CTMO and SDSS source sub-images. Non-transient samples are a pair of sub-images with the same detected source --- one from CTMO and the other from SDSS. Transient samples are a pair of sub-images, one from CTMO containing a source, the other SDSS images containing no source --- only background.

First, we detected sources on the CTMO image. We used the \textit{Source Extraction and Photometry (SEP)} library in Python \citep{bertin_1996} \citep{barbary_2016}. The program detects objects from each image (in this study at 3-$\sigma$ confidence) and provides each of their coordinates as provided by the WCS header solution.

After source extraction, we normalized both images to a common signal level. Each image pair was taken with different instruments, so the first step was to quantify the difference in signal. CTMO images exhibit a much higher resolution than the SDSS ones which had a 3 sigma limit.  The increased depth CTMO images, is possibly due to the their unfiltered nature, whereas the SDSS images were obtained through a \textit{g} band filter.

\begin{figure}
	\centering
	\includegraphics[scale=0.3]{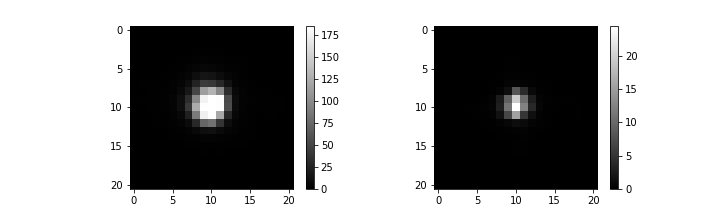} 
	\includegraphics[scale=0.3]{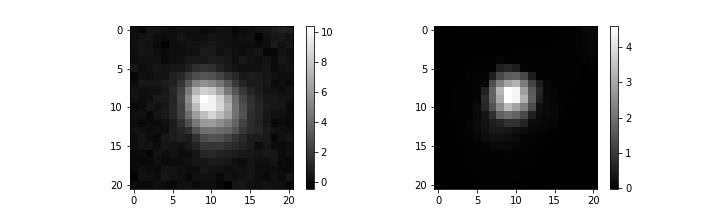}
	\includegraphics[scale=0.3]{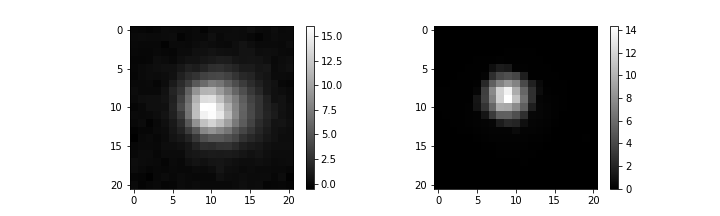}
	\caption{Real non-transient samples}
	\label{real non tr}
\end{figure}

We made sub-images containing a single object from the list of detected sources on each CTMO image. An entire CTMO image is $2048 \times 2048$ pixels and each sub-image was $21 \times 21$ pixels centered on the coordinates of the detected source. Similarly, we made cuts of the aligned SDSS image at each detected source position, giving a pair of cropped images (one from CTMO and one from SDSS showing the same part of the sky). A few examples non-transient samples are shown in Figure \ref{real non tr}.

\begin{figure}
	\centering
	\includegraphics[scale=0.3]{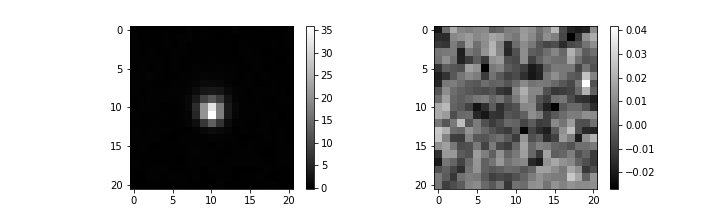}
	\includegraphics[scale=0.3]{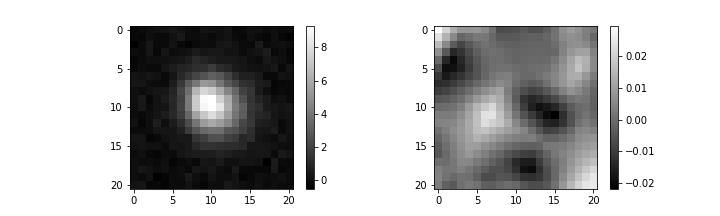} 
	\includegraphics[scale=0.3]{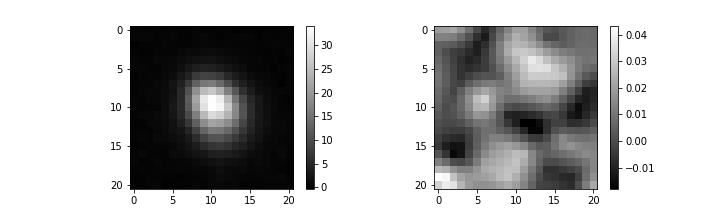} 
	\caption{Real transient samples}
	\label{real tr}
\end{figure}

We did not observe any transients on these images, so we created artificial transient samples. We produced a sub-image containing a single object from the CTMO image (in the way described in the previous paragraph) and chose a spot on the SDSS image where there was only background. A few examples of these transient samples are shown in Figure \ref{real tr}. 

\subsubsection{Simulated data set}

For the training set, we simulated point sources superimposed on a mean background with noise. We fit the parameters of the program to obtain samples that are similar to the samples in the test data set. We set the sample size as an image of $21 \times 21$ pixels. The background variance is generated from a normal distribution with a fixed mean level of zero analog-to-digital units (ADU) and a standard deviation of 0.5 ADU. The profile for each point source is a two-dimensional Gaussian distribution, with different FWHM on the two main axes, and an arbitrary rotation with respect to the $(x, y)$ pixel axes of the image. The FWHM for the major and minor axes are chosen randomly from a uniform distribution. The orientation of the Gaussian profile with respect to the image axis is also selected uniformly over the unit circle.

For the source simulation, it is important to decide which image corresponds to the CTMO and SDSS images. The CTMO sources are brighter and larger in size. We set the amplitude of the brighter source to $(35 \pm 10)$ ADU and FWHM to $(5 \pm 1.5)$ pixels. For the dimmer source, we set the amplitude to $(5 \pm 15)$ ADU and FWHM to $(0.5 \pm 1.5)$ pixels.
	
\begin{figure}
	\centering
	\includegraphics[scale=0.3]{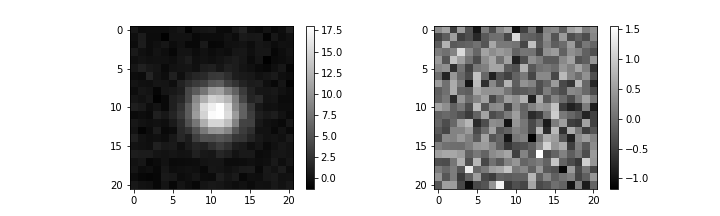}
	\includegraphics[scale=0.3]{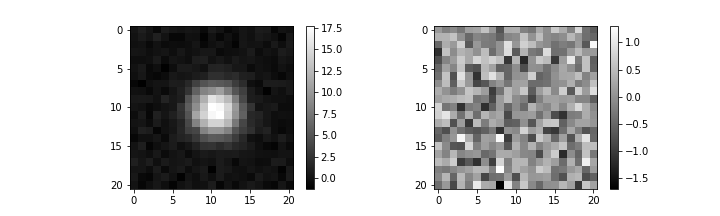}
	\includegraphics[scale=0.3]{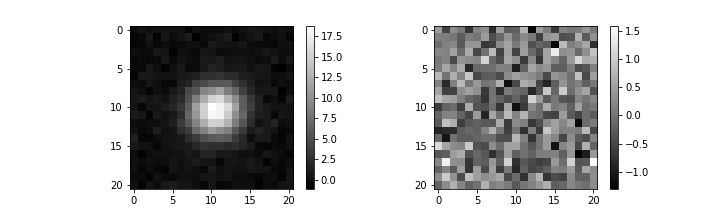}
	\caption{Simulated transient samples}
	\label{sim tr}
\end{figure}

\begin{figure}
    \centering
	\includegraphics[scale=0.3]{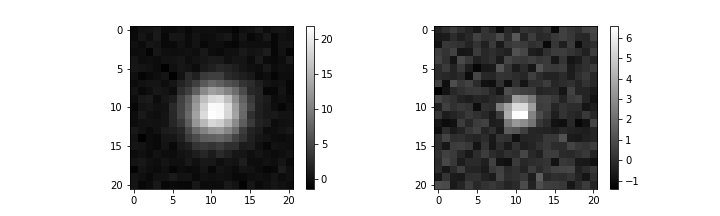} 
	\includegraphics[scale=0.3]{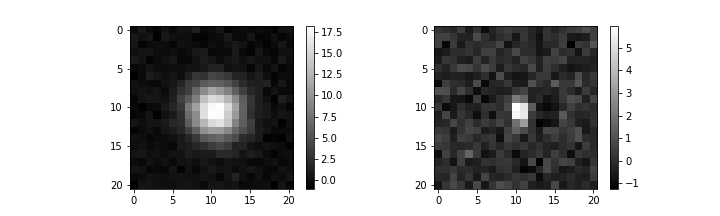} 
	\includegraphics[scale=0.3]{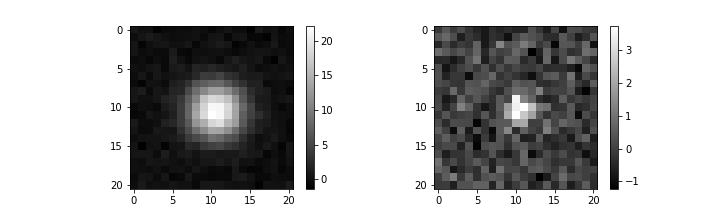} 
	\caption{Simulated non-transient samples}
	\label{non sim tr}
\end{figure}

A sample consists of a pair of small images and labels indicating whether the pair is a transient (label is ``1'') or not (label is ``0''). For non-transient samples, both images contain a simulated object. In this case, on each simulated pair, one source simulates a source expected on CTMO images, while the other simulates SDSS image sources. For transient samples, only one image contains a simulated object, whereas the other contains only simulated background. During the simulation, we chose the likelihood of generating a point source on the background to be 0.5, meaning that 50\% of the samples are transient samples while the rest are non-transient samples. Examples of simulated transient and non-transient samples are shown in Figure \ref{sim tr} and \ref{non sim tr}.

\subsection{Building the Neural Network Models}
\label{sec:cnnarch}

We built two ANN models which we tasked to classify if an image contained a transient. One model uses convolutional layers, which are particularly useful for image analysis \citep{lecun_1990, lecun_1998} (the CNN model). The other model uses dense layers, which are the basic structure of ANNs \citep{mcculloch_1943} (the DLN model). The training process in ML requires fitting a large quantity of free parameters to the model, and, therefore, a large amount of training sample data. Since data containing real transients are scarce, we used simulated samples in the training phase and data collected from real images in a final testing phase. The performance measures we report are from the testing phase. We tested how both models predicted the existence of transients using test image data from CTMO and reference images from SDSS. To download and analyze SDSS images we used the \textit{Astropy} package \citep{robitaille_2013, pricewhelan_2018}. We explain how we generated the training samples and the testing samples in Chapter \ref{sec:datasets}.

We created two types of networks with different topologies -- one a CNN and the other a dense layer network. We trained both networks on the simulated data set and tested them on the test data set. We used the Keras library \citep{chollet_2015} with TensorFlow backend \citep{abadi_2015} to construct the models and scikit-learn libraries \citep{pedregosa_2011} to evaluate prediction of the models.

\subsubsection{Convolutional model with single multi-layer input}

\begin{figure}
    \centering
    \includegraphics[scale=0.3]{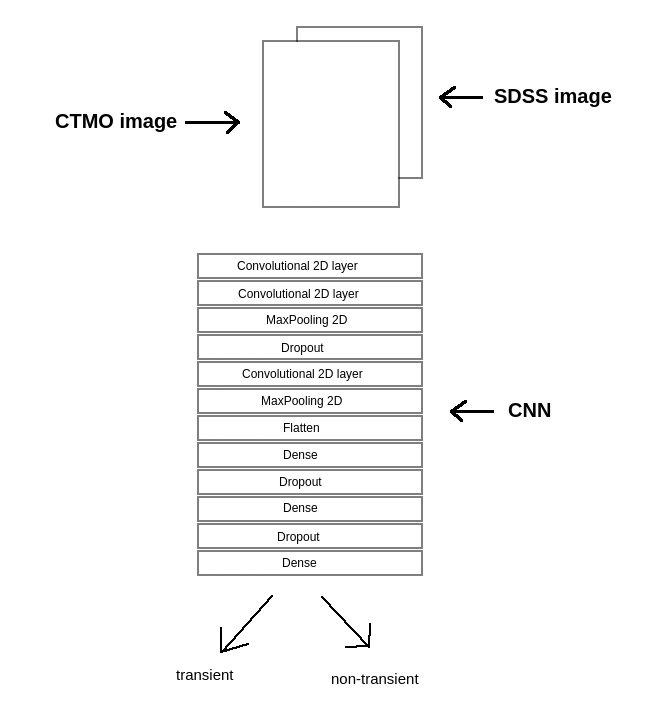} 
    \caption{Schema of the CNN model}
    \label{fig:CNN}
 \end{figure}

We built and tested the first model using convolutional layers, hence it is considered a convolutional model. For this task, we built the network using the sequential model in Keras. As input the model takes one image with two channels -- one channel accepts the CTMO image and the other accepts the SDSS image. The model is a binary classifier -- as an output it returns either ``1'' (a transient sample) or ``0'' (a non-transient sample). The network structure is shown in Figure \ref{fig:CNN}. The number of parameters in each layer and additional properties like the activation function are shown in Table \ref{cnn model summary}. The total number of parameters of the CNN is 1475.

\begin{table}
 	\centering
 	\begin{tabular}{|c|c|c|}\hline
 	\textbf{Layer} & \textbf{Number of parameters} & \textbf{Properties}  \\ \hline
 	Convolutional2D & 190 & AF = \textit{relu} \\ \hline
 	Convolutional2D & 455 & AF = \textit{relu} \\ \hline
 	MaxPooling2D & 0 & pool size = (3, 3) \\ \hline
    Dropout & 0 & 0.25 \\ \hline
    Convolutional2D & 138 & AF = \textit{relu} \\ \hline
 	MaxPooling2D & 0 & pool size = (2, 2) \\ \hline
 	Flatten & 0 & \\ \hline
 	Dense & 40 & AF = \textit{relu} \\ \hline
 	Dropout & 0 & 0.5 \\ \hline
    Dense & 550 & AF = \textit{relu} \\ \hline
 	Dropout & 0 & 0.3 \\ \hline
 	Dense & 102 & AF = \textit{softmax} \\ \hline
 	\end{tabular}
  	\caption{A summary of the CNN model parameters. AF stands for ``activation function''. The \textit{relu} function applies a rectified linear unit activation function. The \textit{softmax} function converts a real vector to a vector of categorical probabilities.}
 	\label{cnn model summary}
 \end{table}

\subsubsection{Dense model with double input}

\begin{figure}
    \centering
    \includegraphics[scale=0.3]{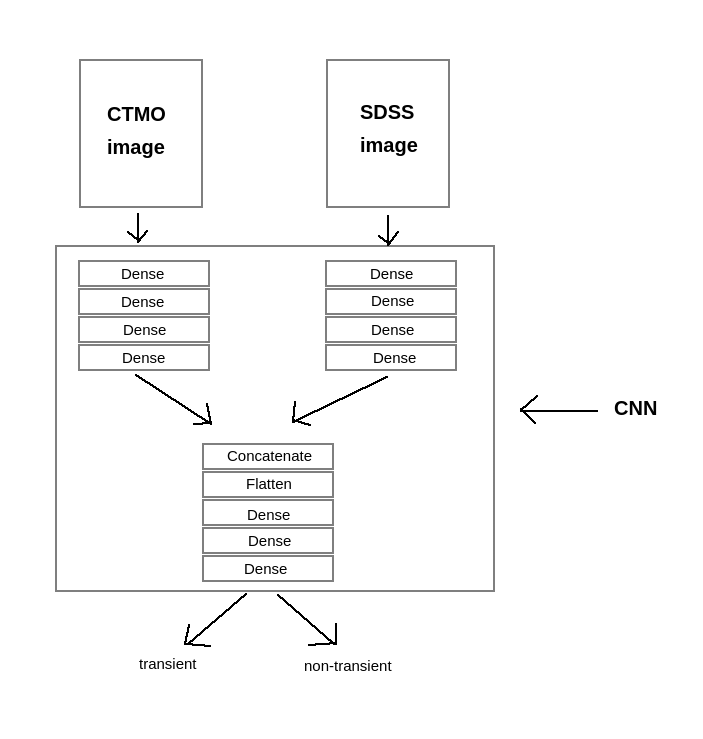} 
    \caption{Schema of the DLN model}
    \label{NN dense}
\end{figure}

In the second model, we use primarily dense layers. As input the model takes two images separately and then combines them. We built network using functional model in Keras. The structure of the network is shown in Figure \ref{NN dense}. The number of parameters in each layer and some additional properties are shown in Table \ref{nn modelsummary}. The total number of parameters of this model is 37594, considerably more than the previous model.

\begin{table}
    \centering
    \begin{tabular}{|c|c|c|}\hline
    \textbf{Layer} & \textbf{Number of parameters} & \textbf{Properties}  \\ \hline
    Input layer 1 & 0 & \\ \hline
    Input layer 2 & 0 & \\ \hline
    Dense input1 & 1408 & 64, AF = \textit{relu} \\ \hline
    Dense input2 & 1408 & 64, AF = \textit{relu} \\ \hline
    Dense input1 & 2080 & 32, AF = \textit{relu} \\ \hline
    Dense input2 & 2080 & 32, AF = \textit{relu} \\ \hline
    Dense input1 & 264 & 8, AF = \textit{relu} \\ \hline
    Dense input2 & 264 & 8, AF = \textit{relu} \\ \hline
    Dense input1 & 36 & 4, AF = \textit{relu} \\ \hline
    Dense input2 & 36 & 4, AF = \textit{relu} \\ \hline
    Concatenate & 0 & \\ \hline
    Flatten & 0 & \\ \hline
    Dense & 21632 & 128, AF = \textit{relu} \\ \hline
    Dense & 8256 & 64, AF = \textit{relu} \\ \hline
    Dense & 130 & 2, AF = \textit{softmax} \\ \hline
    \end{tabular}
    \caption{A summary of the DLN model parameters.}
    \label{nn modelsummary}
\end{table}

\subsection{Validation and Test Metrics}
\label{sec:metricresults}
\begin{figure}
	\centering
	\begin{tabular}{cc}
		\includegraphics[scale=0.24]{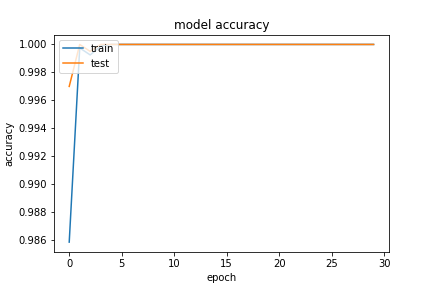} &	\includegraphics[scale=0.24]{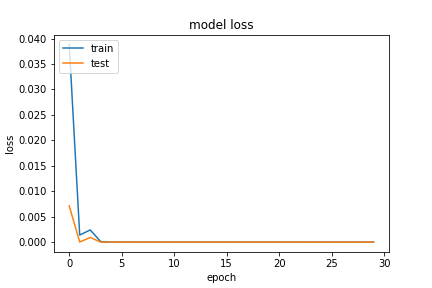}\\
	\end{tabular}
	\caption{The learning process of the CNN model.}
	\label{learningCNN}
\end{figure}

\begin{figure}
	\centering
	\begin{tabular}{cc}
		\includegraphics[scale=0.24]{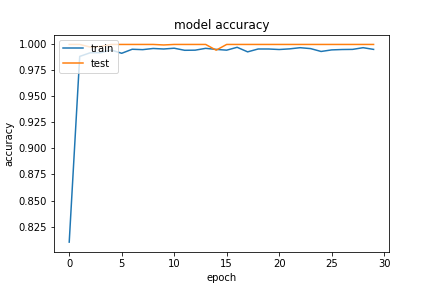} &	\includegraphics[scale=0.24]{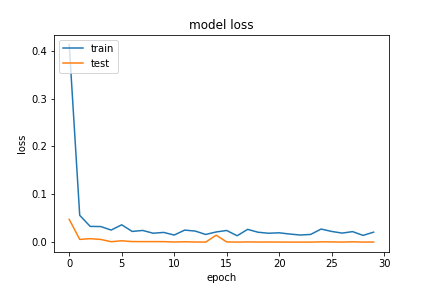}\\
	\end{tabular}
	\caption{The learning process of the DLN model.}
	\label{learningNN}
\end{figure}

We trained both networks using 10000 samples of simulated data. We split the samples into two subset: 8000 samples to train the network and 2000 samples to validate the results. We trained the CNN and DLN models in 30 epochs using the Adam optimizer, and we evaluated the performance of the training with the accuracy metric. The resulting accuracy reflects a compromise between achieving the best results and avoiding an overfitting of the network. The training process is shown in Figures \ref{learningCNN} and \ref{learningNN}.

After training and validation, we calculated the prediction of each model for test data samples. The prediction output is the likelihood of the sample being a transient. A value of one means absolute confidence that the source is a transient, and a value of zero indicates a non-transient source. The confusion matrix is shown in Table \ref{matrices}. The confusion matrix shows how many times the network makes an error and the type of error. The diagonal of the matrix contains the number of correctly classified samples per class, and the off-diagonal elements are the miss-classification for each class. For a two-class system, the off-diagonal elements are the errors of classifying a transient as a non-transient and vice versa.

\begin{table}
    \centering
    \begin{tabular}{|c|c|c|}\hline
        Metric & CNN model score & DLN model score \\ \hline
        Accuracy & 0.989 & 0.969 \\ \hline
        Precision & 0.981 & 0.949 \\ \hline
        Recall & 0.996 & 0.99 \\ \hline
        F1 score & 0.989 & 0.97 \\ \hline
    \end{tabular}
    \caption{Metrics of the CNN and DLN models.}
    \label{metrics}
\end{table}

\begin{table}
    \centering
    \begin{tabular}{|c|c|c|c|}\hline
        & Real / Classified & Non-transient & Transient \\ \hline
        1-model & non-transient & 1653 & 32 \\ \cline{2-4}
        (CNN) & transient & 6 & 1679 \\ \hline
        2-model & non-transient & 1595 & 90 \\ \cline{2-4}
        (DLN) & transient & 13 & 1672 \\ \hline
    \end{tabular}
    \caption{Confusion matrices of the CNN and DLN models.}
    \label{matrices}
\end{table}

The test data consists of 1685 samples of transients and the same amount of non-transients. The CNN model mistakenly classified 32 non-transients as transients and only six transients as non-transients. The dense model made additional errors in non-transient classification. The errors might be caused by the sources having lower statistical significance in the SDSS images in comparison to the CTMO images, so there might be samples in which the SDSS source is of the same order of intensity as the background. The network cannot tell the difference between the dim source and the background, and thus misidentifies these samples as transients.

It is possible to avoid the mistake of false recognition by adding more lower signal-to-noise reference samples into the training data set. Another step could be changing the training data set altogether. If more CTMO data were available, it would be possible to create a training data set from real images in the same way like that for the test data set. Consequently, there would be no need to use simulation data.

\begin{figure}
        \centering
		\includegraphics[scale=0.3]{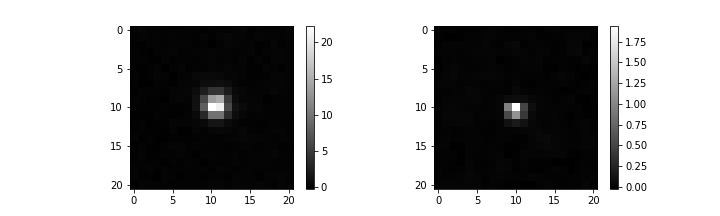} 
		\includegraphics[scale=0.3]{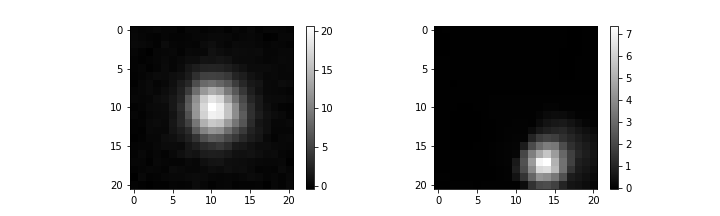}
		\includegraphics[scale=0.3]{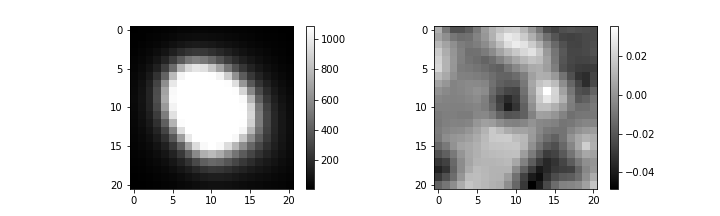}
		\caption{CNN model errors (left column is CTMO and right column is SDSS data).}
		\label{CNN errors}
\end{figure}

\begin{figure}
		\centering
		\includegraphics[scale=0.3]{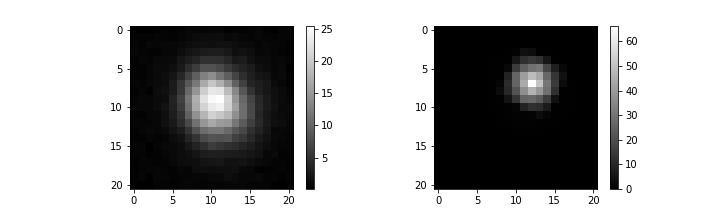} 
		\includegraphics[scale=0.3]{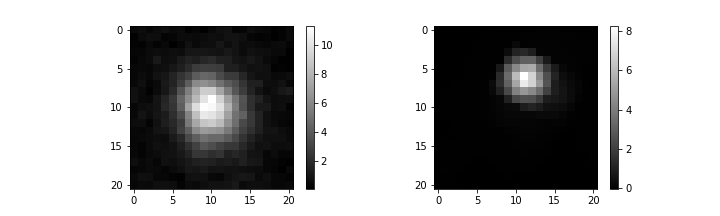}
		\includegraphics[scale=0.3]{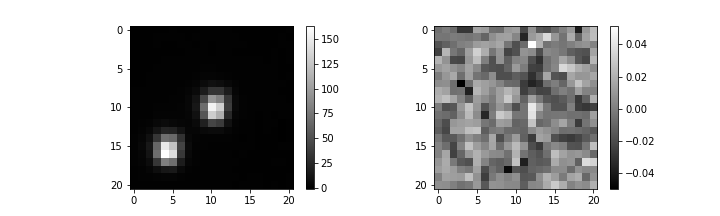}
		\caption{DLN model errors (left column is CTMO and right column is SDSS data).}
		\label{NN errors}
\end{figure}

Regardless, considering the two types of errors, it is preferable to have a non-transient event classified as a transient, not the opposite, because in this case one does not miss any potential transient event. Having a higher miss rate for transients would only cause additional checks for some non-transient cases. Classification error examples are shown in Figures \ref{CNN errors} and \ref{NN errors}. The most common error is produced when the SDSS source is weak. Another type of error is when the CTMO source is bright and large, when it nearly covers the entire sub-image. In one particular case, the network made an error when attempting to identify two sources in one sub-image.

Both models exhibit high accuracy. The accuracy is not 100$\%$ in either case, which means that the networks are not overfitted. The CNN model demonstrated slightly better results than the DLN model, probably caused by the dense layers having many more parameters to train. The performance of the convolutional layers demonstrates that they are generally much better for image analysis. The next step of this project could be to build a model with a double input, such as the DLN model, but using convolutional layers rather than the dense layering.

\section{Comparison of DIA approach and ANN approach on data connected to GW170104}
\label{sec:diaanncompare}

\begin{figure}
    \centering
    \includegraphics[scale=0.1]{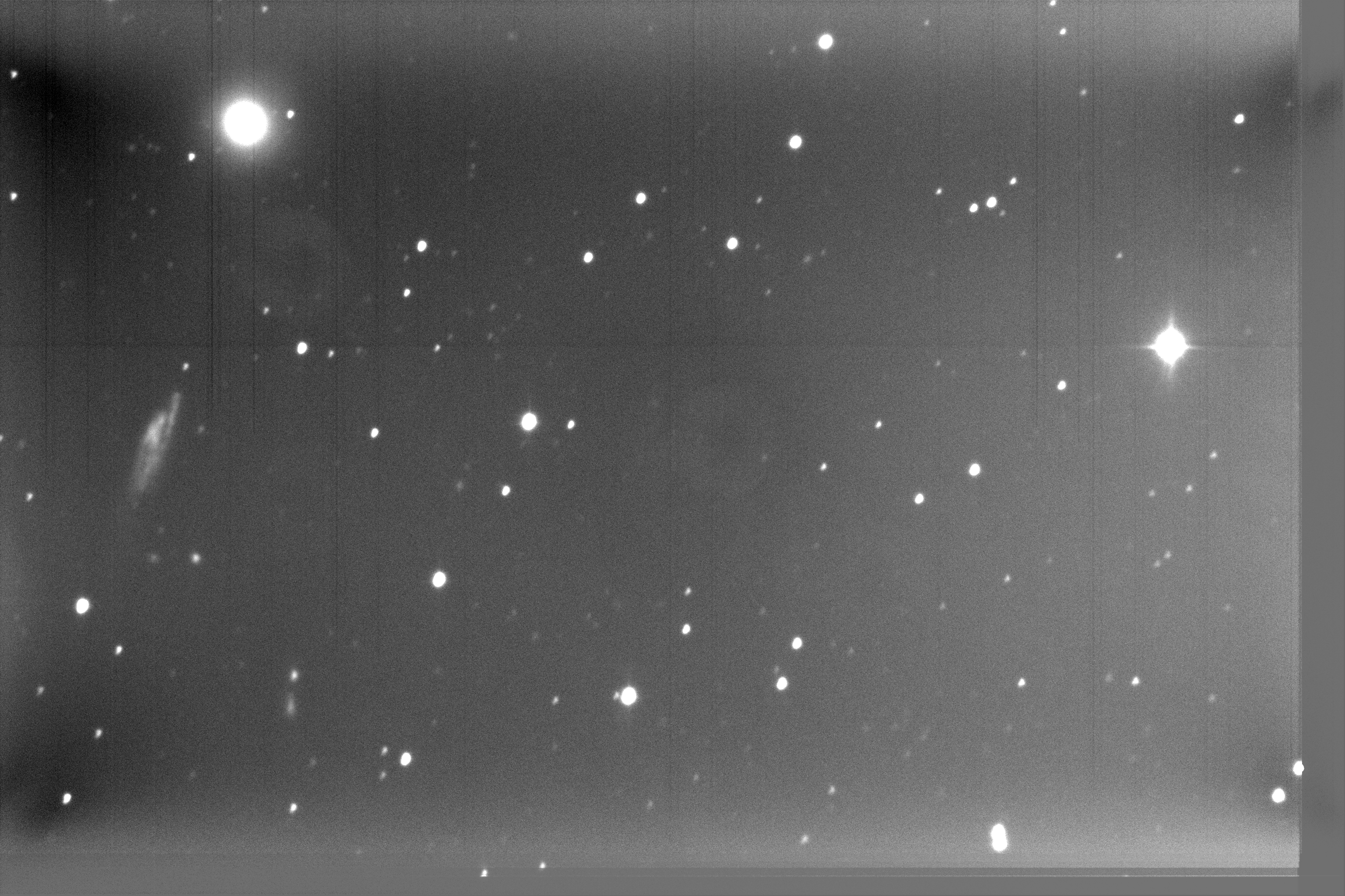} 
    \caption{An image of galaxy ESO 202-009 taken by EABA in January 2017.}
    \label{image o2}
\end{figure}

\begin{figure}
    \centering
    \includegraphics[scale=0.1]{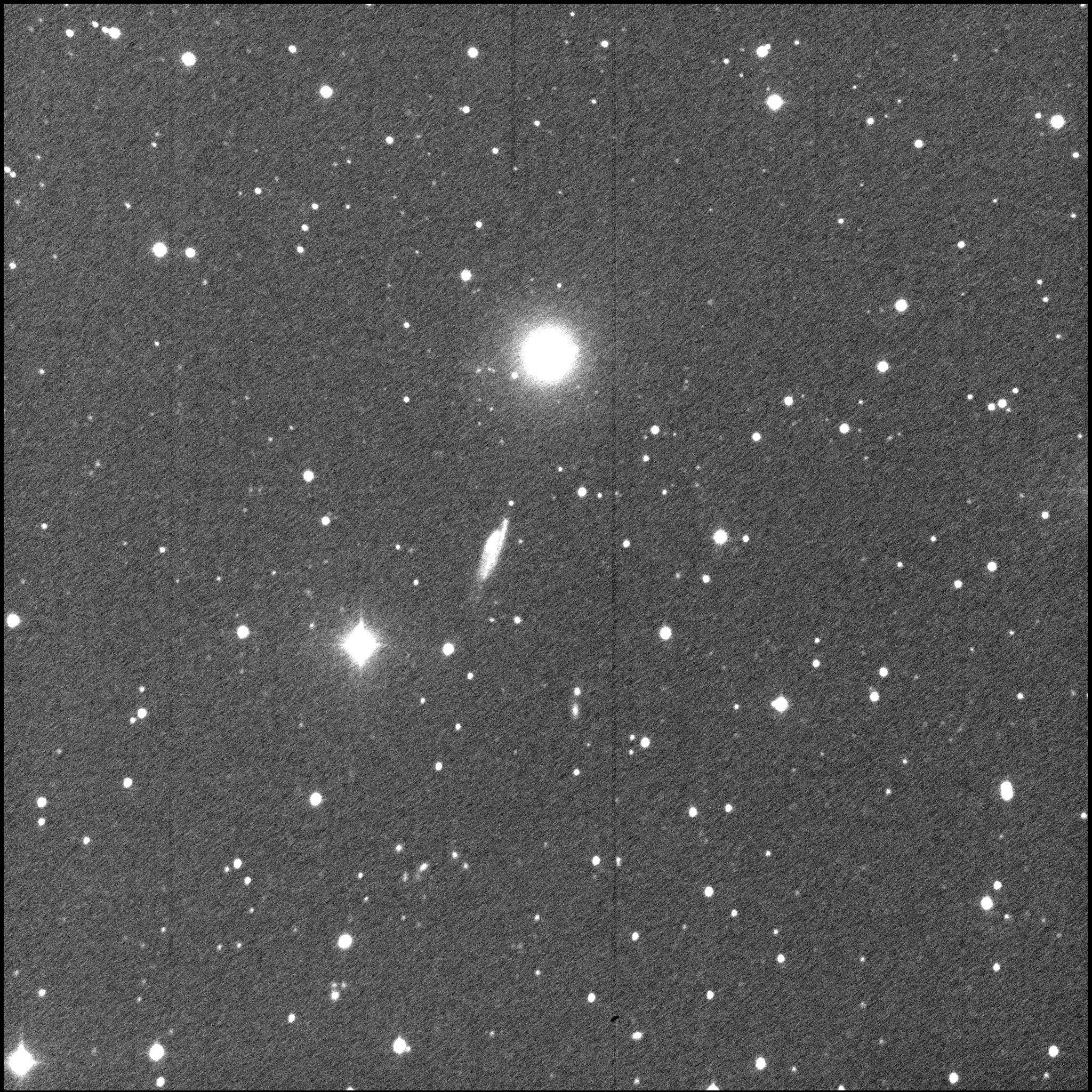} 
    \caption{The reference image of ESO 202-009 taken by EABA in November 2017.}
    \label{ref image o2}
\end{figure}

In this section we would like to present results of comparing DIA approach and ANN approach\footnote{In this a reference image to compare was taken by the same telescope, not an image taken by SDSS.} on search for of optical counterparts connected with GW170104. The initial search for astronomical transients was addressed by \cite{artola_2020}. The authors analyzed images taken by the TOROS Collaboration during the LIGO Scientific Collaboration's second observation run (November 2016 --- August 2017) - O2. TOROS followed up three GW alerts of which two were truly astrophysical: GW170104 and GW170817. In this paper, we only analyze the GW170104 follow-up data. The data for GW170104 were taken by the Estacion Astronomica Bosque Alegre (EABA) in Cordoba, Argentina. TOROS observed the most massive galaxies within the high probability region of localization for the GW events in January 2017, and produced a reference set of the images of the same objects, retrieved later in November 2017. The example of an image set looks like that shown in Figures \ref{image o2} and \ref{ref image o2}.

The transient detection method used by \cite{artola_2020} involved DIA. The main goal of DIA is to transform one image to become compatible with another. The transform involves using a convolutional kernel to reduce the differences in PSFs on both images. The method used by the authors to find and apply the kernel was introduced by \cite{bramich_2008}. Following image transformation, the image is subtracted from the reference to reveal new sources. The DIA method generates a large number of spurious source artifacts (i.e. ``bogus sources''). A typical ratio of real-to-bogus transients is 1:100. A ML algorithm is then used to distinguish between real and bogus sources.

The authors of \cite{artola_2020} generated synthetic "real" sources to create a training set for teaching a ML algorithm to distinguish between real and bogus transients. The method involved repeatedly injecting the profile of a star into an image. Then, they subtracted the images and extracted sources to detect objects on the difference image. Some detected sources were injected objects (i.e. ``real'' transients) and some were artifacts (i.e. ``bogus'' transients). Having samples of real and bogus transients, the authors built and trained a random forest, decision trees, and a support-vector machine --- the best results were obtained by random forest.

Although the problem addressed by \cite{artola_2020} is similar to the one addressed in this paper, the methods are quite different in nature. Models based on DIA distinguish between real and bogus sources collected from a single, difference image. Our method bypasses the subtraction step and, instead, works directly on the target-reference pair of images by focusing on one source at a time and identifying it as a transient or non-transient. Additionally, DIA methods require examples of real and bogus transients to train ML algorithms, while our method requires examples of transients (equivalent to reals) and non-transients. Nevertheless, to compare both methods, we applied the algorithm to the same data used by \cite{artola_2020}.

\begin{figure}
    \centering
    \includegraphics[scale=0.3]{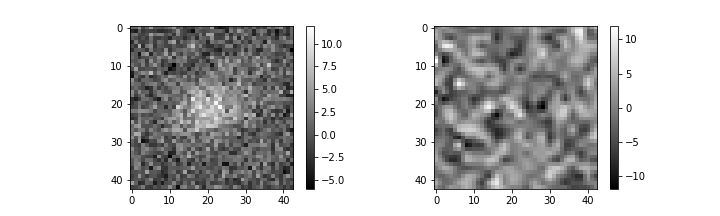} 
    \includegraphics[scale=0.3]{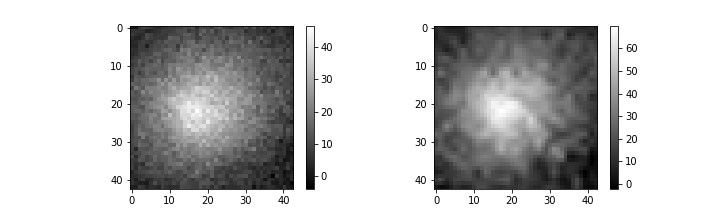} 
	\caption{The top row shows an example of a \textit{real} transient sample and the bottom row shows an example of a \textit{real} non-transient sample.}
	\label{real tr o2}
\end{figure}

We created the training and testing data as follows. We extracted all samples for the test data sets from the original 13 images taken during the GW170104 follow-up event as described in \cite{artola_2020}. The transient samples are the profiles of injected stars on one image and the background on the other image --- they are equivalent to the set of ``real'' transients in the DIA method. The non-transient samples are a pair of thumbnails of the same objects detected by SExtractor in target and reference images. The comparison data set has a total of 3557 samples with labels. An example of transient and non-transient samples is shown in Figure \ref{real tr o2}.

\begin{figure}
    \centering
    \includegraphics[scale=0.3]{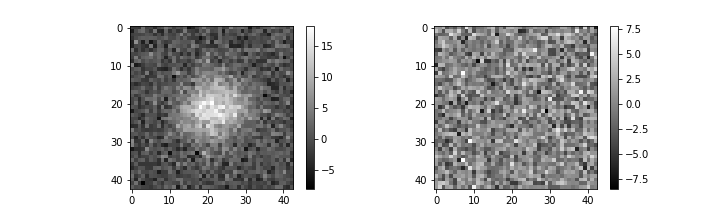}
    \includegraphics[scale=0.3]{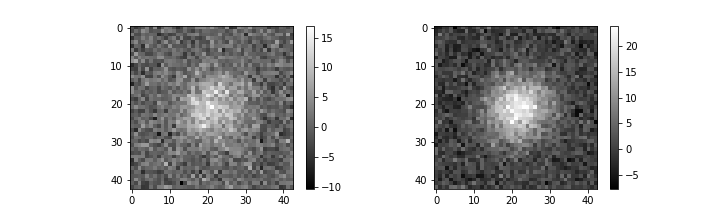} 
    \caption{The top row shows an example of a \textit{simulated} transient sample and the bottom row shows an example of a \textit{simulated} non-transient sample.}
    \label{sim tr o2}
\end{figure}

We retrained the models with different input sizes matching the conditions of \cite{artola_2020}. We simulated a new training data set and we adjusted the background noise level and standard deviation of the simulated training samples to zero and 2.5, respectively, to match those of the test set. Furthermore, we set the amplitude of the simulated sources to an average of 3 ADU and a standard deviation of 10 ADU. The sources are shaped like Gaussian profiles with a $\sigma$ value of (30 $\pm$ 10.5) ADU. In this case, we simulated both sources with the same parameters. creating a total of 10000 samples. Examples of simulated transient and non-transient samples are shown in Figure \ref{sim tr o2}.

The number of parameters to train is different because the size of the sub-image in one sample is bigger ($43 \times 43$ pixels). The total number of parameters of the CNN model is 2195 and for the DLN model is 62938 --- a significant difference. The number of parameters in each model and some additional properties are shown in Tables \ref{cnn modelsummary o2} and \ref{nn modelsummary o2}.

\begin{table}
\centering
\begin{tabular}{|c|c|c|}\hline
        \textbf{Layer} & \textbf{Number of parameters} & \textbf{Properties} \\ \hline
        Convolutional2D & 190 & AF = \textit{relu} \\ \hline
        Convolutional2D & 455 & AF = \textit{relu} \\ \hline
        MaxPooling2D & 0 & pool size = (3, 3) \\ \hline
        Dropout & 0 & 0.25 \\ \hline
        Convolutional2D & 138 & AF = \textit{relu} \\ \hline
        MaxPooling2D & 0 & pool size = (2, 2) \\ \hline
        Flatten & 0 & \\ \hline
        Dense & 760 & AF = \textit{relu} \\ \hline
        Dropout & 0 & 0.5 \\ \hline
        Dense & 550 & AF = \textit{relu} \\ \hline
        Dropout & 0 & 0.3 \\ \hline
        Dense & 102 & AF = \textit{softmax} \\ \hline
    \end{tabular}
    \caption{A summary of the CNN model parameters for O2 data.}
    \label{cnn modelsummary o2}
\end{table}

\begin{table}
    \centering
    \begin{tabular}{|c|c|c|}\hline
        \textbf{Layer} & \textbf{Number of parameters} & \textbf{Properties} \\ \hline
        Input layer 1 & 0 & \\ \hline
        Input layer 2 & 0 & \\ \hline
        Dense input1 & 2816 & 64, AF = \textit{relu} \\ \hline
        Dense input2 & 2816 & 64, AF = \textit{relu} \\ \hline
        Dense input1 & 2080 & 32, AF = \textit{relu} \\ \hline
        Dense input2 & 2080 & 32, AF = \textit{relu} \\ \hline
        Dense input1 & 264 & 8, AF = \textit{relu} \\ \hline
        Dense input2 & 264 & 8, AF = \textit{relu} \\ \hline
        Dense input1 & 36 & 4, AF = \textit{relu} \\ \hline
        Dense input2 & 36 & 4, AF = \textit{relu} \\ \hline
        Concatenate & 0 & \\ \hline
        Flatten & 0 & \\ \hline
        Dense & 44 160 & 128, AF = \textit{relu} \\ \hline
        Dense & 8256 & 64, AF = \textit{relu} \\ \hline
        Dense & 130 & 2, AF = \textit{softmax} \\ \hline
    \end{tabular}
    \caption{A summary of the DLN model for O2 data.}
    \label{nn modelsummary o2}
\end{table}

The confusion matrix for these models is shown in Table \ref{matrices o2}. The main error is, again, in the non-transient classification. In Table \ref{metrics o2}, we compare the metrics for three models: the two networks and the random forest (RF) algorithm tested in \cite{artola_2020}. We cannot treat this comparison as entirely accurate, because the classification problems are inherently different. Regardless, the DLN model obtained the overall best results. 

The main advantage of our method is that it skips the subtraction step, hence it is much quicker and less computationally expensive. Because it skips subtraction, our method can compare images which are significantly different (e.g. taken by different instruments), meaning optical transient could be detected without taking a reference image hours or days later. Hence, our method allows us to detect optical transients with very low-latency. Additionally, our method does not require additional classification between bogus and real transients.  

\begin{table}
    \centering
 	\begin{tabular}{|c|c|c|c|}\hline
 		\textbf{Metric} & \textbf{CNN model score} & \textbf{DLN model score} & \textbf{RF score} \\ \hline
 		Accuracy & 0.91 & 0.918 & 0.89 \\ \hline
 		Precision & 0.856 & 0.866 & 0.92 \\ \hline
 		Recall & 0.993 & 0.997 & 0.86 \\ \hline
 		F1 score & 0.919 & 0.927 & 0.89 \\ \hline
 	\end{tabular}
  	\caption{Metrics of the CNN model, DLN model, and RF algorithm for O2 data.}
 	\label{metrics o2}
\end{table}

\begin{table}
 	\centering
 	\begin{tabular}{|c|c|c|c|}\hline
 		& \textbf{Real/Classified} & \textbf{Non-transient} & \textbf{Transient} \\ \hline
 		Model 1 & non-transient & 1379 & 322 \\ \cline{2-4}
 		(CNN) & transient & 4 & 1842 \\ \hline
 		Model 2 & non-transient & 1403 & 308 \\ \cline{2-4}
 		(Dense) & transient & 12 & 1834 \\ \hline
 	\end{tabular}
  	\caption{Confusion matrices of the CNN model, DLN model, and RF algorithm for O2 data.}
 	\label{matrices o2}
\end{table}

\section{Conclusion}

We have shown that it is possible to detect OTs by comparing images from two different telescopes. Our method opens a new way to search for OTs using reference images from another survey, making it possible to detect an OT in single image taken by a telescope. This feature is especially useful in the fast detection of kilonovae during EM follow-up observations of GW events, and readily adoptable for small observatories to participate in these targets of opportunity.

We tested two neural network models --- one based on CNNs and other based on dense layers. Our models achieved high accuracy (0.989 for the CNN model and 0.969 for the DLN model). The main error in both networks was misidentifying non-transient samples as transients. A reason for false positive detection could be that both images are of different intensity scales (i.e. a given source might have different pixel intensities between target and reference image subsets). There are sample cases for which the object is much weaker in the SDSS image and, therefore, the network sees it as part of background.

We tested both models on data taken by the TOROS Collaboration in follow-up to the GW170104 event. Initially, in order to detect transients in these data, DIA was the primary method, followed by a ML inspection of source-extracted objects on the difference images to distinguish between transients and artifacts. With our method, the models classified whether or not the sample images contained a transient, and they achieved a high accuracy score: 0.91 for the CNN model 0.918 for the DLN model (RF score was 0.89). In this comparative study, the DLN model performed best. 

In order to expand this project, it would be useful to build other models with better efficiency. Models with convolutional layers contain less parameters and, hence, are much easier and quicker to train which will be useful in the analysis of larger images or data sets. A next step could be to combine two different models, like a model with double input but using convolutional layers. Another idea is to use a more advanced network based on CNNs --- generative adversarial networks (GANs) \citep{goodfellow_2016b}. Models using GANs can solve the problem of transforming one image to be similar to another and are, therefore, an alternative to transforming the image convolutional kernel via DIA methods.

The goal of this project is to apply these algorithms to TOROS data and incorporate them into the standard analysis pipeline. The first step is to test the method on TOROS data. We could test if the models detect the real kilonova observed by the TOROS Collaboration in follow-up to GW170817. 

This work is opening a new approach to OT detection, relieving us the need for taking a prior reference image by the same telescope. Instead, we can use an image taken by another image survey as the reference. Our method can help small FOV telescopes to make efficient searches for EM counterparts to GW events. This paper presents a promising beginning for a new class of methods to search for OTs which will be expanded in the future. 

\section*{Acknowledgements}

The TOROS collaboration acknowledges support from Ministerio de Ciencia, Tecnolog\'{\i}a e Innovaci\'on Productiva (MinCyT) and Consejo Nacional de Investigaciones Cient\'{\i}ficas y Tecnol\'ogicas (CONICET) from Argentina, grants from the National Science Foundation of the United States of America, NSF PHYS 1156600 and NSF HRD 1242090, and the government of Salta province in Argentina.

Adam Zadrożny and NCNR is grateful for financial support from MNiSW grant DIR/WK/2018/12 and NCN grant UMO-2017/26/M/ST9/00978.

Katarzyna Wardęga is grateful for scholarship from The University of Texas Rio Grande Valley for the academic year 2019-2020, during which this project was carried out as part of student exchange between University of Warsaw, Faculty of Physics and The University of Texas Rio Grande Valley, Faculty of Physics.

The authors thank Lucas Macri for his helpful comments to the manuscript.

Funding for SDSS-III has been provided by the Alfred P. Sloan Foundation, the Participating Institutions, the National Science Foundation, and the U.S. Department of Energy Office of Science. The SDSS-III web site is http://www.sdss3.org/.

SDSS-III is managed by the Astrophysical Research Consortium for the Participating Institutions of the SDSS-III Collaboration including the University of Arizona, the Brazilian Participation Group, Brookhaven National Laboratory, Carnegie Mellon University, University of Florida, the French Participation Group, the German Participation Group, Harvard University, the Instituto de Astrofisica de Canarias, the Michigan State/Notre Dame/JINA Participation Group, Johns Hopkins University, Lawrence Berkeley National Laboratory, Max Planck Institute for Astrophysics, Max Planck Institute for Extraterrestrial Physics, New Mexico State University, New York University, Ohio State University, Pennsylvania State University, University of Portsmouth, Princeton University, the Spanish Participation Group, University of Tokyo, University of Utah, Vanderbilt University, University of Virginia, University of Washington, and Yale University.

This research has made use of the SIMBAD database,
operated at CDS, Strasbourg, France.

\bibliographystyle{mnras}
\bibliography{references} 

\label{lastpage}
\end{document}